# Hot Electron Production Using the Texas Petawatt Laser Irradiating Thick Gold Targets*

Devin Taylor, Edison Liang, Taylor Clarke, Alexander Henderson, Petr Chaguine, Xin Wang
Rice University

and

Gilliss Dyer, Kristina Serratto, Nathan Riley, Michael Donovan, Todd Ditmire,
University of Texas at Austin

**Abstract**

We present data for relativistic hot electron production by the Texas Petawatt Laser irradiating solid Au targets with thickness between 1 and 4 mm. The experiment was performed at the short focus target chamber TC1 in July 2011, with intensities on the order of several x$10^{19}$W.cm$^{-2}$ and laser energies around 50 J. We discuss the design of an electron-positron magnetic spectrometer to record the lepton energy spectra ejected from the Au targets and present a deconvolution algorithm to extract the lepton energy spectra. We measured hot electron spectra out to ~ 50 MeV, which show a narrow peak around 10 – 20 MeV, plus high energy exponential tail. The hot electron spectral shapes appear significantly different from those reported for other PW lasers. We did not observe direct evidence of positron production above the background.

# 1. Introduction

The study of relativistic hot electron production by ultra-intense laser irradiating solid targets is a timely topic important to many fields, from inertial fusion to laboratory astrophysics. Gamma-ray emission by relativistic electrons is critical to the understanding of many high-energy astrophysical processes including gamma-ray bursts, blazar jets, and pulsar winds. Recently, short-pulse lasers have advanced enough to allow these high-energy astrophysical processes to be studied in the laboratory. In addition to gamma-ray emission, relativistic electrons interacting with high-Z solid targets can produce copious electron-positron pairs in the multi-MeV range [1-3], which also have many laboratory astrophysics applications [7].

The recently commissioned Texas Petwatt Laser (TPW) at UT Austin is one of the world's most intense 100-J class short-pulse lasers [13]. In July 2011, we used the TPW to irradiate thick (1-4 mm) gold targets at the newly completed short-focus target chamber TC1 to study hot electron, gamma-ray and positron production. Even though no positron was convincingly detected due to the high background, this experiment allowed us to determine the background levels to improve our spectrometer designs for later positron experiments. This report focuses on the hot electron data and methodology. Gamma-ray data analysis is in progress and will be reported in a separate paper.

## 1.1 Hot Electron and Pair Production

When an ultra-intense laser strikes a solid target, superthermal "hot" electrons are produced with characteristic energy approximated by [1,9]: $E_{hot} = [(1 + I\lambda^2/1.4\times10^{18})^{1/2} - 1] mc^2$
where I is the laser intensity in W.cm$^{-2}$ and $\lambda$ is the laser wavelength in microns. Up to 30-50% of laser energy can be converted into hot electron energy [9,14]. If the incident laser intensity is such that $E_{hot} > 2mc^2$), these hot electrons can then pair-produce inside a high-Z target [1]. Experimentally, the emergent hot electron spectrum is often quite complicated and depends on details of the target (Z, thickness, density etc) and laser properties (intensity, contrast, polarization, duration, incident angle, focal spot size etc, see [14] for review). Some experiments measuring mainly the low energy (< few MeV) spectrum show that the hot electron temperature $T_{hot}$ may be more accurately approximated by $T_{hot} \sim (I\lambda^2)^{0.34}$, consistent with the Beg scaling model [11], while other results seem to favor the ponderomotive scaling $T_{hot} \sim (I\lambda^2)^{0.5}$ above [1,9]. As we see in Sec.5, the TPW hot electron spectrum cannot be described by simple exponentials.

The process of electron-positron pair production from hot electrons interacting with a high-Z nucleus can occur through two channels: the Trident process or the Bethe-Heitler (BH) process [3,10]. In the Trident process, electrons create pairs by directly interacting with the nucleus. This process dominates for thin foils <10's of microns [1]. We are using thick targets (1- 4 mm) in our experiment, thus BH dominates: hot electrons first emit bremsstrahlung gamma-rays which then interact with the nucleus to create an electron-positron pair [3]. Both processes depend on $Z^2nL$ where n=ion density and L=target thickness. Hence BH scales as $Z^4n^2L^2$, strongly favoring thick Au (Z=79) targets [1]. However, for L > few mm, the pairs cannot escape from the target. Hence the optimal thickness for laser pair creation is around 3 – 4 mm [1].

When the laser-driven plus secondary electrons exit the target, they create a sheath electric field that can accelerate the positrons plus any surface protons. The sheath electric field forms on the order of a few tens of femtoseconds and can be very intense for high-Z targets [12]. Most positrons may travel through this field, gain energy equal to the sheath potential and shift the peak energy of the observed positron spectrum up by several MeV [12]. This field also helps the positrons to form a narrow jet out of the target back [3], along an axis between the laser forward and target normal directions [12]. At the same time, the sheath potential likely reflux some of the low energy electrons back into the target and broaden the electron distribution, thus altering the detected electron spectrum. Detailed modeling of the emergent electron and positron spectra requires using a combination of PIC and particle physics Monte Carlo codes. So far there has been a lack of end-to-end simulations of the emergent electron and positron spectrum from first principles. Hence much more experimental data is critically needed to advance this field.

1.2 Texas Petawatt Laser

The experiment was conducted in TC1 of the Texas Petawatt Laser located on the University of Texas at Austin campus. The laser is based on an optical parametric chirped pulse amplification (OPCPA) design with mixed silicate and phosphorus Nd:glass amplification. Such a design allows a shorter pulse duration and higher intensity on target [6]. The design specifications of the laser give an estimated maximum energy on target of up to 200 J with a pulse duration of 150 fs and spot size of 5 microns [6]. The laser contrast is estimated to be around $10^7$ and perhaps as high as $10^{12}$ [13]. At full power, we expect a peak intensity of $I > 10^{21}$ W/cm$^2$. However, during our experiment, the laser was kept below maximum power to avoid damage to the f/3 focusing optics. Thus we saw an energy range of 40 – 60 J, and a longer pulse duration on the order of 200 – 300 fs, focused to a maximum intensity of $< 10^{20}$ W/cm$^2$. We were assigned one week of shot time in July 2011, and carried out 14 shots.

## 2. Experimental Set-up

The goal of the experiment is to measure hot electron, gamma-ray and positron production from an ultra-intense laser incident on thick Au targets. Au (Z = 79) is chosen because of its high-Z and high density. Motivated by theory and previous experiments [1-3], targets of 1 – 4 mm thickness were used.

To measure the spectra of laser-produced electrons and positrons, a magnetic spectrometer is designed and fabricated as the principle diagnostics device. The magnetic spectrometer was placed in the target normal position as indicated in Figure 2.1. The spectrometer's designed energy range is ~1-50 MeV. To complement this high-energy spectrometer, a low energy weak-field spectrometer is borrowed from UT-Austin to cover the energy range < 6 MeV. The data recording mechanism of both spectrometers is chosen to be phosphorus image plates (Fuji BAS SR2040). The plates do not require development and are reusable. In addition, the image plates can be read quickly via a computerized scanner and their response to deposited electrons has been well studied and calibrated [2,5]. These plates offer the benefit of a quick read time without complicated electronics inside the chamber. However, the data is wiped by visible light and degrades over time. Therefore the spectrometer case must be light-tight and image plates must be digitally recorded within 90 minutes or risk data loss [2].

Figure 2.1: The target chamber set-up for the TPW experimental run in July 2011 (the radial lines are in 10 degree increments about laser forward direction). The yellow central box indicates the orientation of our targets at 17 degrees from laser forward. (1) The high energy e+/e- spectrometer was placed at a variable distance of 9-22 cm at target normal. (2) The low energy e+/e- spectrometer was first placed at (a) 4 degrees outside laser forward and later placed at (b) the front of the target. (3) The filter-stack gamma ray spectrometers: (a) looking at the high energy spectrometer positron side and (b) various other locations looking directly at the target.

## 3. The e+e- Spectrometer

3.1 Design

Our primary diagnostic is a magnetic e+e- spectrometer fabricated at Rice University. The magnetic field necessary to give the desired energy range (1-50 MeV) is ~0.6T. Since image plates are sensitive to x-rays, a major effort of the design is to reduce the x-ray background via optimal shielding. The spectrometer consists of three components: the outer case, the inner spectrometer, and the shielding. Because the response of the spectrometer could not be determined prior to the experiment beyond Monte Carlo simulations, the spectrometer is designed to be adaptable. The outer case is designed to be light tight, because the image plates used in the spectrometer are wiped by visible light.

The inner spectrometer consists of two 2" wide x 6" long neodymium-iron-boride (Nd-Fe-B) magnets separated by a distance of 1.4 cm to achieve peak magnetic field strength of approximately 0.6T in the gap. The magnets are separated by Fe yokes to contain the magnetic field. so that the electrons will not be significantly diverted from the central axis until they are within the gap. This helps to improve the lower energy resolution of the spectrometer. The downside of using Fe yoke is that Fe fluorescence creates more internal background. Because of the fringe magnetic field geometry at the magnet edges and the width of the magnet, electrons tend to be focused towards the mid-plane of the gap, which enhances the signal to background on the image plates.

Because the image plates are slightly magnetic, they attach to the magnets automatically without any special holder. A thin cavity is etched into the Al siding of the spectrometer to accomodate the plates, which measure 1" x 6" and run the length of the gap. The outer case has room for up to 4" of front shielding. We used alternating layers of 0.5" Pb and 0.5" of Cu with a total thickness of 2'. A 3mm collimating pinhole is bored through the shielding.

Figure 3.1 Magnetic spectrometer with Al case and 3 mm pinhole. The outer case dimensions are 3.5 inches x 4.5 inches x 12 inches. It can measure e+/e- energies from ~ 1 MeV – 50 MeV.

3.2 Calibration

The low energy calibration can be accomplished with a standard radioactive source. $^{90}$Sr was used because the emitted electron energy cutoff is relatively high at 2.28 MeV. To calibrate the high energies, it is necessary to go to an electron beam line. We used the LSU Mary Bird Perkins Cancer Center Elekta clinical beams at Baton Rouge, which has monoenergetic electron beams of ~6–22 MeV, allowing multiple calibration points. Several data points were taken with different beam energies in early 2012 to verify the Monte Carlo simulated energy spectrum based on the 3D-measured B-field map. In Fig.3.2 we show that the LSU electron beam data and the simulated position data are found to agree to better than < 1 mm [8]. Based on such agreement, the Monte Carlo simulated point spread functions are taken to be good approximations of the true spectrometer point spread functions and are thus used in the deconvolution algorithm.

Figure 3.2: Calibrated electron energy spectrum from LSU data points [8] and $^{90}$Sr source.

## 4. Data Analysis

Due to the non-uniform magnetic field in the spectrometer, the deconvolution of the energy spectrum from the recorded images requires a good estimate of the point spread functions of the spectrometer, which are energy dependent. In addition, large backgrounds, especially in the low energy regime, require careful subtraction to extract the data without losing the low energy signal. Once these aspects of the data have been determined, the spectral data can then be deconvolved using a response matrix of the system for some input energy spectra and minimizing the predicted positions against the measured data.

4.1 Background Subtraction

The first step in extracting the energy spectrum of the electron and positron signal is the removal of the background signal. The background signal is determined to be a result of several different processes. Foremost, there is a nearly constant background that results from external gamma-rays striking the case, the magnets and the image plates. The internal background is a result of electron bremsstrahlung, secondary photoelectrons, scattered gamma rays from the pinhole, plus electron reflux onto the Fe yoke. The large background at the front end of the image plates is due to the Fe yoke fluorescence: The low energy electrons are refluxed onto the iron yoke by the magnetic field and in turn generate x-rays and secondary electrons. Other background is due to the electrons striking the magnets. To remove most of the background signal, the background is chosen to be the average of the signal inside the cavity above and below the electron signal. As can be seen in Figure 4.1, the defined background offers reasonable agreement with the electron signal profile. Unfortunately, simulations indicate that the lowest energy point spread functions have a large vertical spread. This would indicate that our method for background subtraction will remove some of the lowest energy electrons. Regardless, this method seems more accurate than removing only the background recorded in the magnet region (pink area), which is the background only due to external radiation.

Figure 4.1: A schematic of the background profile. The pink regions (a) are the magnets and are excluded. The blue regions (b) are the spaces in the cavity between the signal and the magnets that are used to create the background profile. The background is taken as an average of the top and bottom of the cavity between magnets (blue regions (b)).

4.2 Image Deconvolution

With the background removed from the image plate, we must now vertically integrate the signal to get a position spectrum. From the position spectrum, we wish to extract the energy spectrum. To do this, we must first construct a response matrix for the system. The response matrix is constructed from point spread functions generated using the GEANT4 Monte Carlo code from CERN. The point spread functions are created in 0.1 MeV increments to allow for fine energy deconvolution. Sample PSF's are displayed in Figure 4.2. The energy-position spectrum is well matched to the calibration points gathered from the LSU MBPCC electron beam lines [8] and a $^{90}$Sr source.

Figure 4.2: A view of several point spread functions generated from the output data of the GEANT4 Monte Carlo code.

We next deconvolve the shot data using this response matrix. The deconvolution is non-trivial since we have to solve a large matrix equation of the form **Ax = b,** where **A** is our response matrix, **x** is the unknown incident energy spectrum, and **b** is the measured position spectrum. Simply inverting the matrix may not produce desirable or smooth results since the problem is not well posed. It is therefore necessary to verify and improve the inverted spectrum by performing a minimization routine. The minimization routine is used to optimize the solution of $\|\mathbf{Ax} - \mathbf{b}\|^2 = 0$.

Figure 4.3: Example of background subtraction for a 1 mm shot. The shot data is summed vertically (red line) and the data along two strips between the signal and the magnets is taken to be the background (blue line). When no peak is visible as in Fig.(b), the resultant signal is taken as an upper limit.

Taking the position spectrum, and using the Monte Carlo-generated response matrix, we can convert the data from position space to energy space. We present the deconvolved spectrum in 1 MeV energy bins. The error is taken to be the standard deviation of the position spectrum from the IP data. To estimate the error in the final energy spectrum, the input position spectrum is varied in a normal random distribution based on the standard deviation of the data in the position space. We then pass this position spectrum through the response matrix many times to estimate an error for the energy spectrum based on the error from the position spectrum. A sample result is given in Figure 4.4.

Figure 4.4: A sample deconvolved spectrum from the data presented in Figure 4.3a. The deconvolution is done in 1 MeV bins. Left is the spectrum in linear-linear plot. Right is the same spectrum in log-linear plot showing the exponential tail.

4.3 Peak Energy and Effective kT

As we see in Fig.4.4, the typical deconvolved hot electron spectrum can be characterize by two key parameters, the peak energy $E_{pk}$ where the spectrum turns over, and the effective kT of the exponential tail (slope in a log-linear plot). In Sec.5 we study the correlation between these empirical parameters and incident laser intensity. The effective kT is extracted by fitting the high energy tails of the deconvolved spectrum to an equation of the form $N \sim \exp(-E/kT)$ where N is the signal and E is energy. An example of this fitting mechanism can be seen in Figure 4.5.

Figure 4.5: Sample exponential fit to extract effective kT.

## 5. Main Results and Interpretations

After completing the analysis, we were not able to detect any convincing positron signal above the background (Fig.4.3). This is not unexpected since the laser intensity was below $10^{20}$ W/cm$^2$ and the x-ray background was very high, especially at low energies. However, we were able to extract useful hot electron spectra. From the data we can compare $E_{pk}$ and effective kT and their relation to incident laser intensity as noted in Figure 5.1.

In examining the energy spectra, the most surprising feature is the steep turnover of low-energy electrons <10 MeV (see Figure 4.4). Previous experiments carried out at other PW lasers such as Titan [3,11], Vulcan and Omega EP [12], as well as our own PIC simulations, have found much broader electron spectra with more abundant low energy electrons (<10 MeV). In other words, the hot electrons accelerated by TPW have higher average energy than those found in other short-pulse laser experiments or $E_{hot}$ given by the equation in Sec.1.1. This deficit of low energy electrons is curious and could be indicative that in TPW, the hot electrons may be energized by mechanisms other than ponderomotive acceleration [9,14]. In particular, the existence of a narrow peak around 10-20 MeV may be indicative of underdense acceleration mechanisms in the pre-plasma, such as LWFA [15] or reverse sheath acceleration due to electron-ion charge separation [14]. These results remain to be confirmed in future TPW experiments. We see a positive correlation between intensity and peak energy in Figure 5.1a. While the best-fit slope favors $kT\sim I^{0.5}$ instead of $kT\sim I^{0.34}$ scaling, the absolute kT values are higher than those given by the relation in Sec.1.1. We also observe a weak positive correlation between kT and intensity in Figure 5.1b. Finally, in Figure 5.1c, we see a tight correlation between $E_{pk}$ and kT. More data is needed to confirm these trends, and the physics behind such correlations remains to be understood from first principles.

Figure 5.1: The correlations between measured shot intensity, peak energy, and effective temperature kT for 1mm Au targets irradiated by the TPW. Straight lines are best power-law fits.

## 6. Discussions

The lack of any clear positron signal could be due to a combination of several factors: (a) our laser energy per shot is only ~ 50 J. This is much lower than the Titan and Omega-EP shots by Chen et al [3,11,12]. (b) Lower laser energy means fewer exiting hot electrons and lower sheath electric field, which renders emergent positrons to have too low an energy to be observed above the background. (c) Our background may be too higher compared with Titan and Omega-EP shots due to insufficient shielding. Future TPW experiments should improve on all of these.

The deficit of low energy electrons and the narrow electron peak make the TPW electron distribution different from other reported hot electron spectra. This may be caused by the unique properties of the TPW laser, especially its short pulse (< 0.3 ps). If confirmed by future experiments, this bodes well for potential applications of such electron beams to medical therapy and other narrow-band electron applications.

In Fig.2.1 we showed other detectors in the experiment besides the primary e+e- spectrometer. Unfortunately the low-energy e+e- spectrometer 2 did not return useful data due to insufficient shielding. The filter-stack gamma-ray spectrometer 3 did obtain bremsstrahlung spectra (< 3 MeV) emitted by the target front surface (position 3b at various angles). They are much softer than the hot electrons emitted at the target back, as expected. So they cannot be used for cross-calibration of the hot electron temperature. Gamma-ray signals from position 3b are too weak to confirm the presence of positrons. In addition, we have covered the outside surface of the target chamber with dozens of gamma-ray dosimeters. The total gamma-ray dose agrees with the Monte Carlo simulated dose using the hot electron spectra of Sec.4, to within a factor 2. Gamma-ray data analyses and calibration are still in progress and will be reported in a future paper.

**Acknowledgement**

This work was supported by the DOE grant DE-SC-000-1481 and Rice University Faculty Initiative Fund.

# Appendix A

# Table of Shot Parameters

| Shot Number | Target | Position | Image Plate Fade Time | Peak Laser Intensity |
|---|---|---|---|---|
| 2336 | 3mm Au | 20cm | 32 min | $7.0 \times 10^{19} W/cm^2$ |
| 2340 | 3mm Al | 20cm | 29 min | $2.5 \times 10^{19} W/cm^2$ |
| 2341 | 3mm Au | 29cm | 32 min | $2.1 \times 10^{19} W/cm^2$ |
| 2342 | 3mm Au | 29cm | 34 min | $4.3 \times 10^{19} W/cm^2$ |
| 2349 | 2mm Au | 18cm | 28 min | $100 TW^*$ |
| 2353 | 2mm Au | 18cm | 44 min | $2.4 \times 10^{19} W/cm^2$ |
| 2354 | 4mm Au | 16cm | 25 min | $200 TW^*$ |
| 2356 | 2mm Au | 16cm | 20 min | $177 TW^*$ |
| 2357 | 1mm Au | 16cm | 23 min | $1.6 \times 10^{19} W/cm^2$ |
| 2364 | 1mm Au | 16cm | 19 min | $6.1 \times 10^{19} W/cm^2$ |
| 2365 | 1mm Au | 16cm | 21 min | $1.4 \times 10^{19} W/cm^2$ |
| 2366 | 1mm Au | 16cm | 20 min | $1.6 \times 10^{19} W/cm^2$ |
| 2368 | 1mm Au | 16cm | 18 min | $2.6 \times 10^{19} W/cm^2$ |
| 2369 | 1mm Au | 16cm | 16 min | $8.1 \times 10^{19} W/cm^2$ |

*Farfield of laser off screen, approximate measurement of intensity not possible.

Table A.1: |Target and spectrometer properties for each shot as well as observed intensity data.

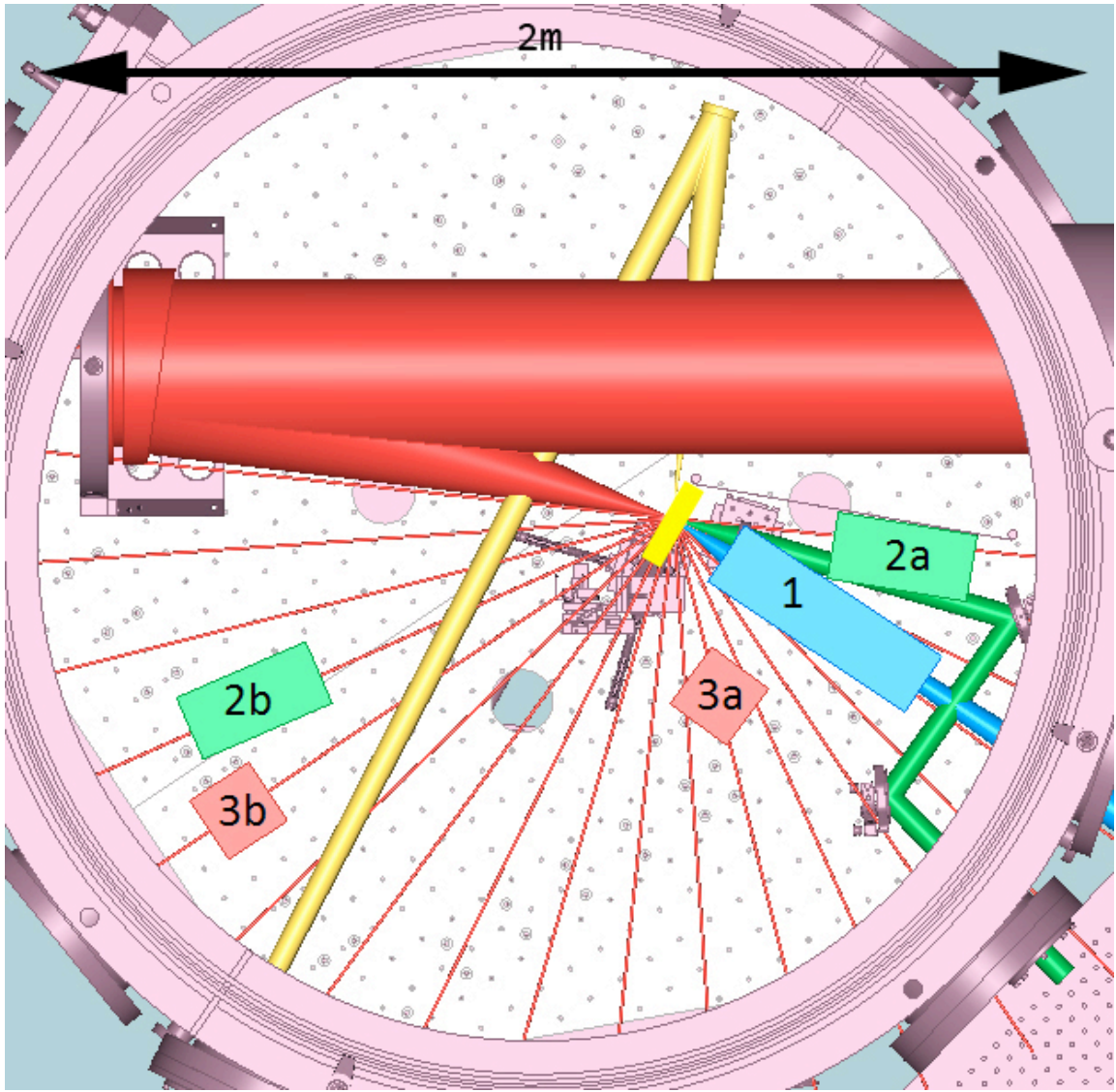

Fig.2.1

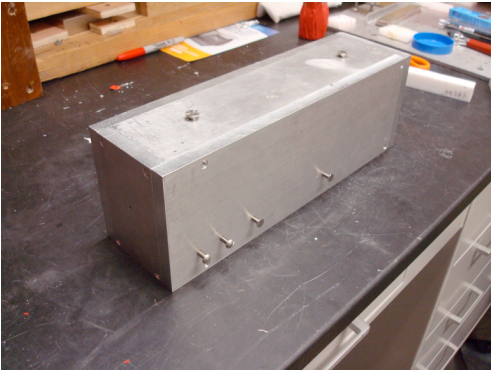

Fig.3.1

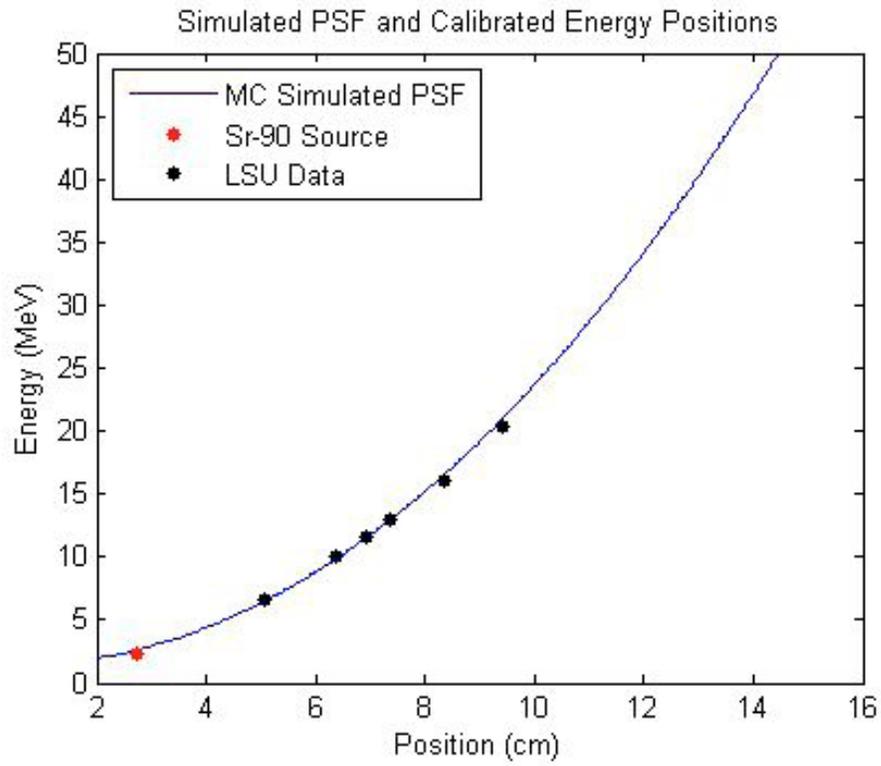

Fig.3.2

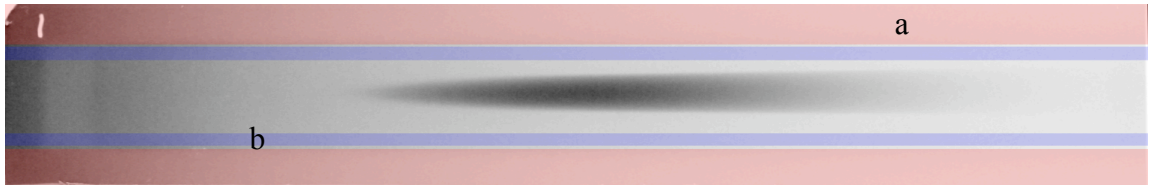

Fig.4.1

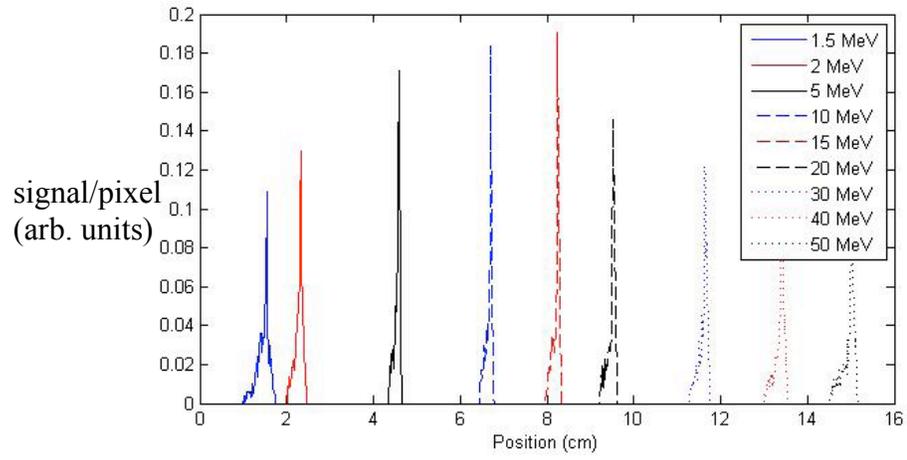

Fig.4.2

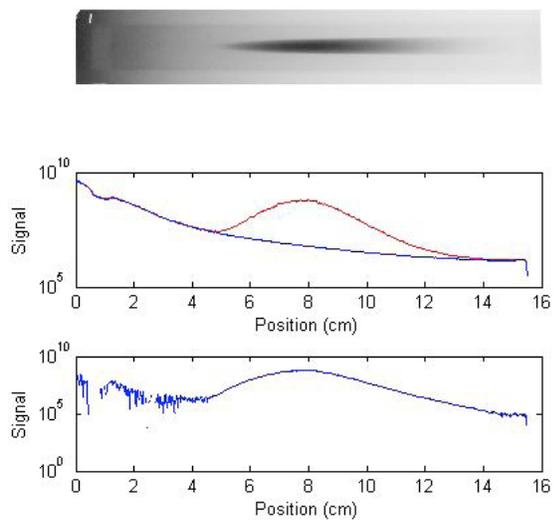 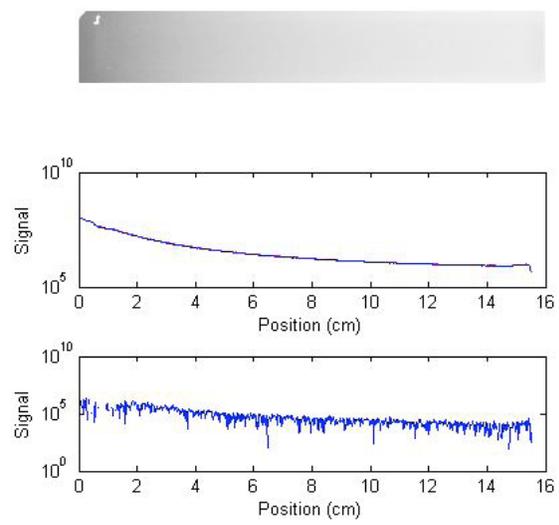

(a) Electron Data Sample  (b) Positron Data Sample

Fig.4.3

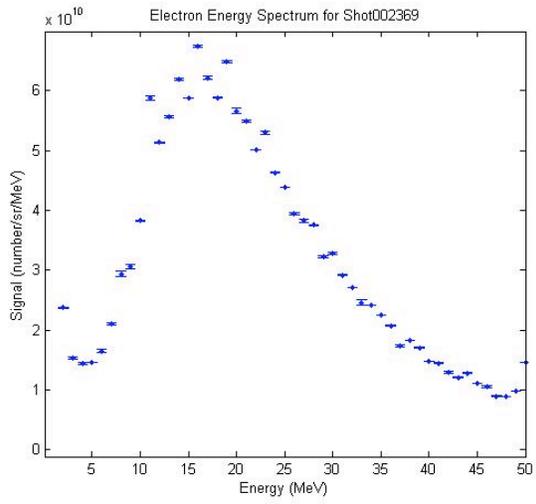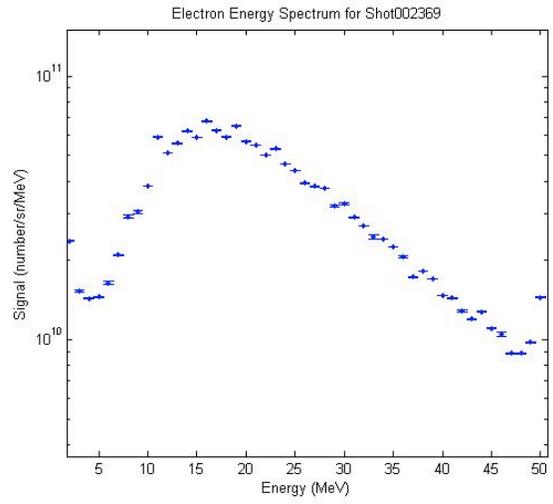

Fig.4.4

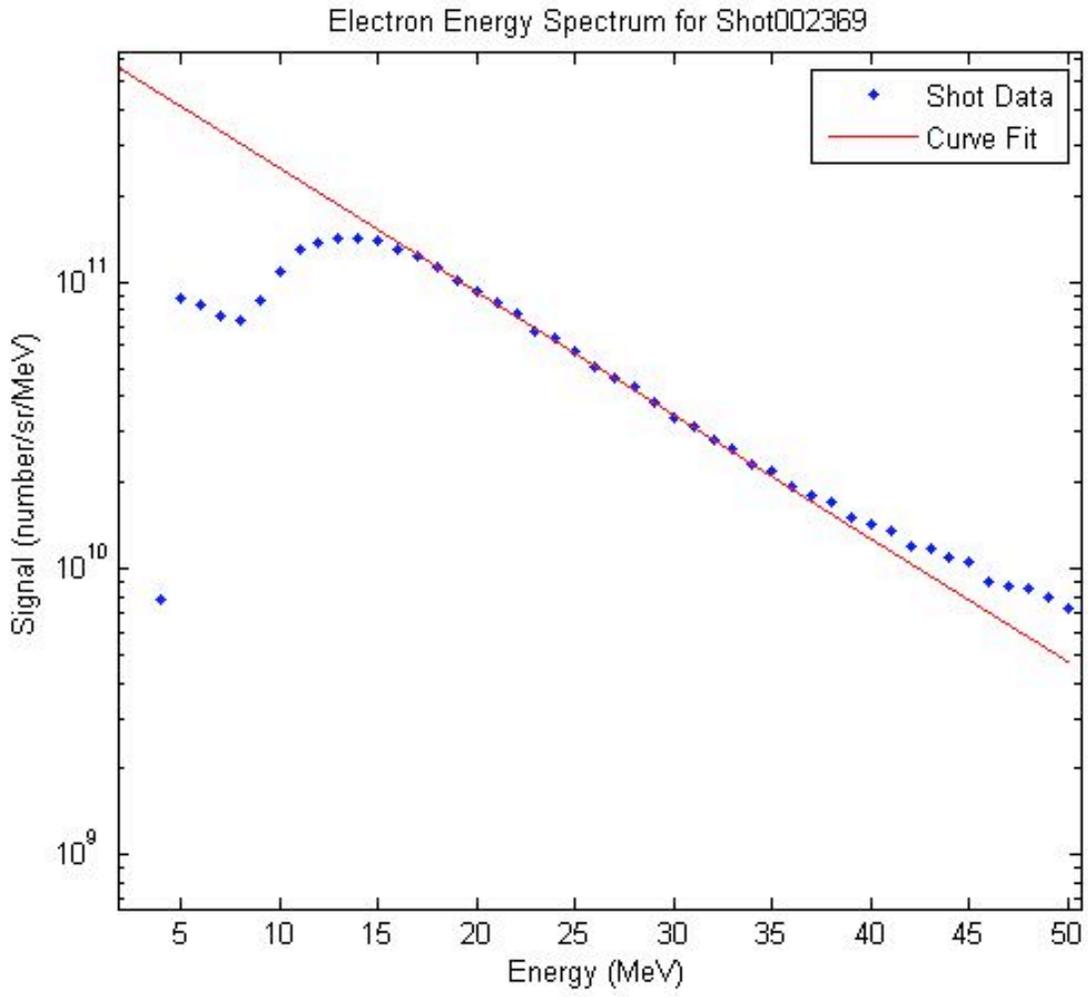

Fig.4.5

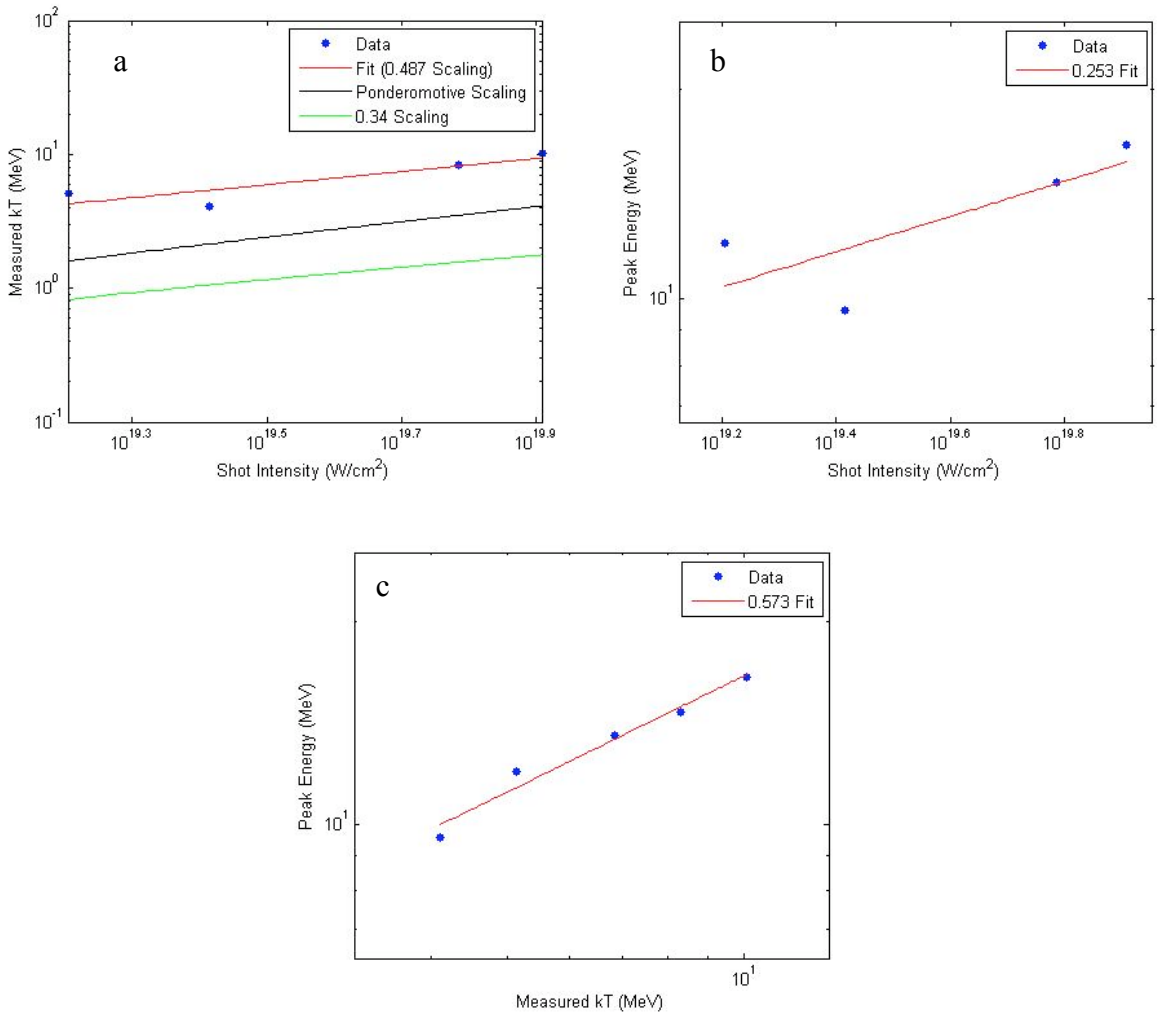

Fig.5.1